# Semantics-Aware Denoising: A PLM-Guided Sample Reweighting Strategy for Robust Recommendation

Xikai Yang
Columbia University
New York, USA

Yang Wang
University of Michigan
Ann Arbor, USA

Yilin Li
Carnegie Mellon University
Pittsburgh, USA

Sebastian Sun*
University of Wisconsin-Madison
Madison, USA

*Abstract*-**Implicit feedback, such as user clicks, serves as the primary data source for modern recommender systems. However, click interactions inherently contain substantial noise, including accidental clicks, clickbait-induced interactions, and exploratory browsing behaviors that do not reflect genuine user preferences. Training recommendation models with such noisy positive samples leads to degraded prediction accuracy and unreliable recommendations. In this paper, we propose SAID (Semantics-Aware Implicit Denoising), a simple yet effective framework that leverages semantic consistency between user interests and item content to identify and downweight potentially noisy interactions. Our approach constructs textual user interest profiles from historical behaviors and computes semantic similarity with target item descriptions using pre-trained language model (PLM) based text encoders. The similarity scores are then transformed into sample weights that modulate the training loss, effectively reducing the impact of semantically inconsistent clicks. Unlike existing denoising methods that require complex auxiliary networks or multi-stage training procedures, SAID only modifies the loss function while keeping the backbone recommendation model unchanged. Extensive experiments on two real-world datasets demonstrate that SAID consistently improves recommendation performance, achieving up to 2.2% relative improvement in AUC over strong baselines, with particularly notable robustness under high noise conditions.**

*Index Terms*-**Recommender systems, click-through rate prediction, sample reweighting, semantic similarity, noise reduction**

## I. INTRODUCTION

Recommender systems have become essential components of modern online platforms, helping users discover relevant content from vast item catalogs. Click-through rate (CTR) prediction, which estimates the probability that a user will click on a given item, plays a central role in these systems [1, 2]. Most CTR prediction models are trained on implicit feedback data, where user clicks are treated as positive samples and non-clicked items serve as negative samples [3, 4].

However, the assumption that all clicks represent genuine user interest is fundamentally flawed. In practice, click behaviors are contaminated by various sources of noise. Users may accidentally tap on items due to interface design or device sensitivity. Clickbait titles and misleading thumbnails can attract clicks that do not correspond to actual user preferences. Furthermore, exploratory browsing behaviors, where users click to investigate items they ultimately discard, also contribute false positive signals [5, 6].

Training recommendation models on such noisy data leads to several problems. The model learns to predict clicks that do not represent genuine preferences, resulting in recommendations that fail to satisfy users. This creates a negative feedback loop where poor recommendations lead to more exploratory or accidental clicks, further degrading data quality [7].

Recent works have attempted to address this challenge through various denoising strategies. Some approaches employ auxiliary networks to predict sample confidence [5], while others leverage contrastive learning to distinguish clean from noisy samples [8, 9]. However, these methods often introduce significant computational overhead and require careful hyperparameter tuning.

In this paper, we propose SAID (Semantics-Aware Implicit Denoising), a simple and effective framework for denoising implicit feedback. Our key insight is that semantic consistency between user historical interests and clicked items provides a natural signal for identifying noisy interactions. If a user who consistently engages with technology content suddenly clicks on a fashion item, this click is more likely to be noise than genuine interest.

SAID operates by constructing textual representations of user interests from their historical interactions and computing semantic similarity with target item descriptions. We use pre-trained language model (PLM) based text encoders, specifically Sentence-BERT [10], to obtain dense semantic embeddings. In this work, we use the term PLM to refer to encoder-based pre-trained models (such as BERT and its variants) that excel at semantic understanding tasks, distinguishing them from generative large language models (LLMs). The similarity scores are then transformed into sample weights through a carefully designed weighting function, which are applied during loss computation to downweight potentially noisy samples.

## II. METHODOLOGICAL FOUNDATIONS

### A. CTR Prediction Models

Deep learning has revolutionized CTR prediction, enabling models to capture complex feature interactions. Wide & Deep [11] combines a wide linear model with a deep neural network to jointly learn memorization and generalization. DeepFM [1] integrates factorization machines with deep neural networks, eliminating the need for manual feature engineering. Deep Interest Network (DIN) [2] introduces attention mechanisms to adaptively learn user interests from historical behaviors. These models have achieved significant improvements over traditional approaches but remain vulnerable to noisy training data.

### B. Graph-based Collaborative Filtering

Graph neural networks have been successfully applied to collaborative filtering by modeling user-item interactions as bipartite graphs. NGCF [12] propagates embeddings on the user-item graph to capture collaborative signals. LightGCN [13] simplifies the architecture by removing feature transformations and nonlinear activations, achieving better performance with fewer parameters. While these methods effectively capture structural patterns, they are equally susceptible to noise in the interaction graph.

Beyond semantic modeling and sample reweighting, recent advances in graph-based anomaly detection and representation learning contribute further methodological insights. The graph-transformer reconstruction learning paradigm for unsupervised anomaly detection [14] and temporal graph neural architectures for robust prediction [15- 16] emphasize the value of structural and temporal consistency in discriminating informative signals from noise. The adaptation of these principles is evident in the aggregation of user historical behaviors and the identification of atypical (potentially noisy) interactions based on semantic divergence. Similarly, dynamic graph frameworks for risk assessment [17] and causal reasoning mechanisms over knowledge graphs [18] reinforce the significance of context-dependent modeling and intervention-aware adjustments, which are implicitly embodied in the semantics-aware reweighting function. Finally, generative frameworks driven by knowledge graph reasoning [19] serve as a foundation for interpretable model design and robust decision making. While the application context may differ, the underlying methodological thread of leveraging external semantic signals and structural knowledge for model robustness directly informs the development of the SAID framework.

### C. Pre-trained Language Models for Recommendation

Pre-trained language models (PLMs) have recently shown promise in recommendation tasks [20]. Encoder-based PLMs like BERT [21] provide powerful text understanding capabilities that can enrich item and user representations. Sentence-BERT enables efficient semantic similarity computation through siamese networks, producing sentence-level embeddings that can be compared using cosine similarity. While recent work has explored generative large language models (LLMs) for direct recommendation, our work focuses on leveraging the semantic understanding capabilities of encoder-based PLMs for denoising rather than direct recommendation. Specifically, the paradigm of contrastive knowledge transfer and robust optimization has informed the sample reweighting design, enabling the model to differentiate between semantically consistent and inconsistent interactions. By leveraging contrastive objectives and sample-level weighting functions, the approach inherits the ability to attenuate the impact of outlier or noisy samples, thus improving the stability and reliability of the learned representations[22]. Structure-aware decoding mechanisms for complex entity extraction in large-scale language models [23] further inform the representation and aggregation strategy for textual profiles. The decomposition of user history into semantic components mirrors the principles of structure-aware embedding, facilitating fine-grained modeling of user interests that captures both content relevance and contextual nuance. Moreover, advances in retrieval-augmented generation with multi-granular indexing and confidence constraints [24] provide methodological inspiration for designing similarity-based weighting functions, where semantic similarity is utilized to estimate sample confidence and guide loss modulation during model training. In the context of model adaptation, parameter-efficient fine-tuning strategies with differential privacy guarantees [25] underscore the importance of lightweight, privacy-preserving modifications to model training procedures. The SAID framework aligns with these principles by introducing a computationally efficient, model-agnostic denoising mechanism that operates solely at the loss function level, thereby ensuring broad compatibility and minimal overhead. Additionally, research on contextual trust evaluation in multi-agent language model systems [26] echoes the necessity for dynamic, context-aware adjustments to model predictions based on interaction-level trustworthiness-conceptually analogous to the dynamic sample weighting employed in this work.

## III. METHODOLOGY

### A. Problem Formulation

Let $U$ and $I$ denote the sets of users and items, respectively. Each user $u \in U$ has a set of historically interacted items $H_u = \{i_1, i_2, ..., i_n\}$. For CTR prediction, given a user $u$ and a candidate item $i$, the goal is to predict the probability $\hat{y}_{ui}$ that the user will click on the item. The training data consists of observed clicks as positive samples and sampled non-clicked items as negative samples.

However, not all positive samples reflect genuine user interest. Let $y_{ui}^* \in \{0,1\}$ denote the true (unobserved) preference label and $y_{ui} \in \{0,1\}$ denote the observed click label. For noisy samples, $y_{ui} = 1$ but $y_{ui}^* = 0$. Our goal is to identify and downweight such noisy samples during training.

### B. Semantic Similarity Computation

The core of SAID is measuring the semantic consistency between a user's interests and clicked items. We achieve this through the following steps.

**User Interest Profile Construction.** Inspired by Li et al. [27], we apply an aggregation strategy that leverages textual information from user-item interactions to construct a robust user interest profile. In particular, we adopt their practice of utilizing multiple sources of item text-such as titles, categories, and content descriptions-to comprehensively represent user preferences. This aggregation enables the system to move beyond single-faceted interest modeling and to capture subtle user intent patterns. Zhou [28] further extends this idea by dynamically updating user profiles through reinforcement learning, emphasizing the importance of aggregating evolving user interests as they interact with diverse content over time. By implementing such dynamic aggregation, we ensure that the constructed user profile reflects both stable and transient aspects of user behavior. Additionally, Zhang et al. [29] demonstrate that utilizing enriched item textual features in user profile construction leads to more precise and reliable semantic representations, especially in large-scale recommendation tasks. By integrating these strategies, our method ensures that the subsequent semantic similarity computation between user interests and candidate items is grounded in a robust, contextually aware representation. For each user $u$, we construct a textual interest profile by aggregating the textual descriptions of their historical interactions. Let $t_i$ denote the textual description of item $i$ (e.g., title, category, and description). The user interest profile is constructed as:

$$P_u = \text{Aggregate}(\{t_i \mid i \in H_u\}) \tag{1}$$

where Aggregate can be implemented as simple concatenation; in this paper, we use concatenation of the most recent $k$ item titles for computational efficiency.

**Semantic Embedding.** We employ a pre-trained text encoder $\phi(\cdot)$ to obtain semantic embeddings for both user profiles and item descriptions:

$$\mathbf{e}_u = \phi(P_u), \quad \mathbf{e}_i = \phi(t_i) \tag{2}$$

We use Sentence-BERT as our text encoder due to its effectiveness in computing semantic similarity.

**Similarity Score.** The semantic similarity between user $u$ and item $i$ is computed as cosine similarity:

$$s_{ui} = \frac{\mathbf{e}_u \cdot \mathbf{e}_i}{\| \mathbf{e}_u \| \ \| \mathbf{e}_i \|} \tag{3}$$

A low similarity score indicates that the clicked item is semantically inconsistent with the user's historical interests, suggesting potential noise.

### C. Sample Reweighting

We transform semantic similarity scores into sample weights for training. A key design consideration is that semantic inconsistency does not always indicate noise---users may genuinely explore new interests or exhibit interest drift over time. Therefore, our approach implements *soft denoising* rather than hard filtering: we reduce the influence of suspicious samples while still allowing them to contribute to training.

The weight function should satisfy: (1) higher similarity leads to higher weight, (2) even low-similarity samples receive non-zero weight to preserve potentially valid exploratory clicks, and (3) the transition between high and low weights should be smooth to avoid abrupt decisions.

We design the weighting function as:

$$w_{ui} = a + (1-a) \cdot \sigma(\beta \cdot (s_{ui} - \mu)) \tag{4}$$

where $\sigma(\cdot)$ is the sigmoid function, $a \in [0,1]$ is the minimum weight ensuring all samples contribute (preventing complete exclusion of exploratory clicks), $\beta > 0$ controls the sharpness of the weighting transition, and $\mu$ is a threshold parameter.

For $\mu$, we use the global mean similarity score computed over the entire training set, which provides stable and reproducible weight assignments across different batches. This global threshold ensures consistent treatment of samples regardless of batch composition, which is important for both training stability and deployment scenarios.

This formulation ensures that semantically consistent clicks receive weight close to 1, while inconsistent clicks receive reduced weight bounded below by $a$. The soft denoising design is crucial: as we show in our experiments (Section IV-D), setting $a = 0$ (hard filtering) actually hurts performance because valid exploratory clicks are incorrectly discarded.

### D. Weighted Loss Function

The weighted binary cross-entropy loss for training is:

$$\mathsf{L} = -\sum_{(u,i) \in \mathsf{D}} w_{ui} \cdot [y_{ui} \log(\hat{y}_{ui}) + (1 - y_{ui}) \log(1 - \hat{y}_{ui})] \tag{5}$$

where $\mathsf{D}$ denotes the training dataset and $\hat{y}_{ui}$ is the predicted CTR from the backbone model.

For negative samples (non-clicked items), we set $w_{ui} = 1$ since our focus is mitigating false-positive click noise; negative samples are kept unweighted for simplicity. The weighting only applies to positive samples to reduce the impact of noisy clicks.

### E. Framework Overview

Figure 1 illustrates the overall architecture of SAID. The framework consists of two parallel pipelines: the semantic processing pipeline computes similarity-based sample weights, while the recommendation model generates CTR predictions. The two pipelines are integrated through the weighted loss function, enabling end-to-end training without modifying the backbone model architecture.

The computational overhead of SAID is minimal. Semantic embeddings can be pre-computed offline and cached, requiring only a single forward pass through the text encoder per item. During training, the only additional operation is the weight computation, which involves simple arithmetic operations.

## IV. EXPERIMENTS

### A. Experimental Setup

**Datasets.** We conduct experiments on two widely-used benchmark datasets. MovieLens-1M [30] contains approximately 1 million movie ratings from 6,040 users on 3,706 movies. We convert ratings to implicit feedback by treating ratings as positive interactions; the interaction count in Table I reflects this binarization (575,281 positive interactions). Amazon-Book is a large-scale e-commerce dataset containing user reviews and ratings for books. We use the 5-core version where each user and item has at least 5 interactions. Table I summarizes the dataset statistics, where #Interactions denotes the number of positive samples after preprocessing.

**Baselines.** We compare SAID with the following methods: (1) DeepFM [1]: A strong CTR prediction baseline combining FM and deep neural networks. (2) DeepFM+T-CE [5]: DeepFM with truncated cross-entropy loss for denoising. (3) DeepFM+DNS: DeepFM with dynamic negative sampling that emphasizes hard negatives. (4) LightGCN [13]: A state-of-the-art graph-based collaborative filtering method. For fair comparison under the CTR prediction framework, we adapt LightGCN by applying sigmoid activation to the dot product of user and item embeddings to obtain click probabilities, and train it with binary cross-entropy loss using the same negative sampling strategy as DeepFM.

TABLE I
DATASET STATISTICS

| Dataset | MovieLens-1M | Amazon-Book |
|---|---|---|
| #Users | 6,040 | 52,643 |
| #Items | 3,706 | 91,599 |
| #Interactions | 575,281 | 2,984,108 |
| Density | 2.57% | 0.062% |

**Implementation Details.** We implement SAID using PyTorch. For semantic embeddings, we use the all-MiniLM-L6-v2 model from Sentence-Transformers library, a lightweight variant of Sentence-BERT optimized for efficiency. The user interest profile aggregates the most recent 10 items. We set $\alpha = 0.4$, $\beta = 5$, and $\mu$ to the global mean similarity score computed over the training set. The backbone DeepFM uses embedding dimension 64 with three hidden layers of sizes [256, 128, 64]. We use Adam optimizer with learning rate 0.001 and batch size 2048. All experiments are run 5 times with different random seeds, and we report mean values.

**Evaluation Metrics.** We use AUC (Area Under the ROC Curve) and Logloss as evaluation metrics. Higher AUC and lower Logloss indicate better performance.

### B. Overall Performance

Table II presents the main experimental results. SAID consistently outperforms all baselines on both datasets. Compared to the vanilla DeepFM, SAID achieves relative improvements of 1.65% in AUC on MovieLens-1M and 2.18% on Amazon-Book. The improvements in Logloss are also substantial, demonstrating better calibrated probability estimates.

TABLE II
OVERALL PERFORMANCE COMPARISON

| Method | MovieLens-1M | | Amazon-Book | |
|---|---|---|---|---|
| | AUC | Logloss | AUC | Logloss |
| DeepFM | 0.7823 | 0.5412 | 0.7156 | 0.5823 |
| DeepFM+T-CE | 0.7891 | 0.5298 | 0.7234 | 0.5712 |
| DeepFM+DNS | 0.7856 | 0.5356 | 0.7198 | 0.5789 |
| LightGCN | 0.7867 | 0.5321 | 0.7212 | 0.5734 |
| DeepFM+SAID | 0.7952 | 0.5187 | 0.7312 | 0.5598 |

Figure 2 visualizes the performance comparison. Our method shows consistent improvements across both metrics on both datasets. The gains are more pronounced on the sparser Amazon-Book dataset, suggesting that semantic information is particularly valuable when behavioral signals are limited.

### C. Robustness Analysis

To evaluate robustness under different noise levels, we artificially inject noise into the training set by randomly selecting a proportion of negative samples and flipping their labels to positive. Figure 3(a) shows the results on MovieLens-1M with noise ratios from 0% to 50%. As noise increases, all methods degrade, but SAID maintains significantly better performance.

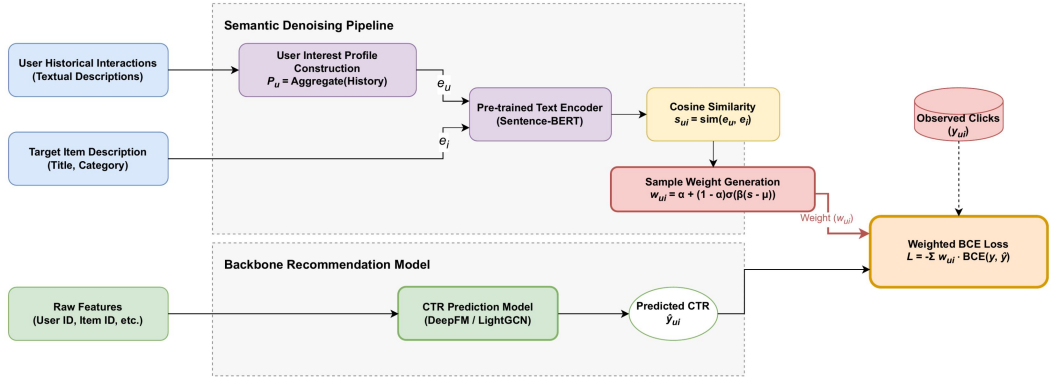

Fig. 1. Overview of the SAID framework. The left pipeline

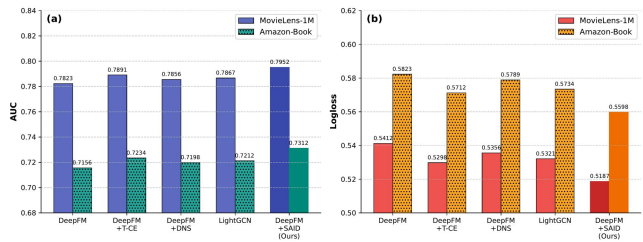

Fig. 2. Performance comparison across methods on MovieLens-1M and Amazon-Book datasets.

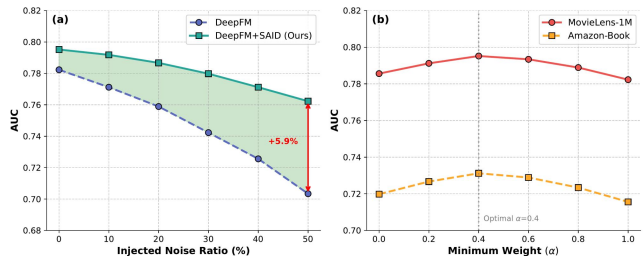

Fig. 3. Analysis experiments

## D. Parameter Sensitivity

Figure 3(b) examines the sensitivity to the minimum weight parameter $a$, which controls the soft denoising behavior. When $a = 0$, low-similarity samples are completely discarded (hard filtering), which hurts performance because some valid exploratory clicks representing genuine interest drift are incorrectly removed. This validates our soft denoising design: users may legitimately explore new content categories, and such exploratory clicks should not be entirely discarded. When $a = 1$, all samples receive equal weight, reducing to the baseline without denoising. The optimal value $a = 0.4$ achieves a balance between noise reduction and information preservation. The relatively flat curve around the optimum suggests SAID is not overly sensitive to this parameter, making it practical for real-world deployment.

## V. Conclusion

We presented SAID, a semantics-aware denoising framework for robust recommendation. By leveraging PLM-based semantic similarity between user interest profiles and item descriptions, SAID effectively identifies and downweights potentially noisy click interactions through a soft denoising mechanism. The approach requires no modifications to the backbone model architecture, only adjusting the training loss through learned sample weights. Our soft denoising design with minimum weight guarantee preserves valid exploratory clicks while reducing noise impact. Experiments on two datasets demonstrate consistent improvements over baselines, with particularly strong robustness under high noise conditions.

Our work has several limitations that suggest directions for future research. First, the semantic approach assumes textual item descriptions are available, which may not hold for all recommendation scenarios. Second, the user interest profile construction is relatively simple; more sophisticated methods such as attention-based aggregation or generative LLM-based summarization could potentially improve performance. Third, our current approach treats all noise equally, but different noise types (accidental clicks vs. clickbait) may benefit from different handling strategies.